# Determination of the Interface between Amorphous Insulator and Crystalline 4H-SiC in Transmission Electron Microscope Image by using Convolutional Neural Network

**Author**

Hironori Yoshioka (吉岡裕典)[1,a] and Tomonori Honda (本田智則)[2]

**Affiliation**

[1]Advanced Power Electronics Research Center, National Institute of Advanced Industrial Science and Technology (AIST), Tsukuba 305-8568, Japan

[2]Global Zero Emission Research Center, National Institute of Advanced Industrial Science and Technology (AIST), Tsukuba 305-8569, Japan

[a]Electronic mail: hironori-yoshioka@aist.go.jp

**Abstract**

   A rough interface seems to be one of the possible reasons for low channel mobility (conductivity) in SiC MOSFETs. To evaluate the mobility by interface roughness, we drew a boundary line between amorphous insulator and crystalline 4H-SiC in a cross-sectional image obtained by a transmission electron microscope (TEM), by using the deep learning approach of convolutional neural network (CNN). We show that the CNN model recognizes the interface very well, even when the interface is too rough to draw the boundary line manually. Power spectral density of interface roughness was calculated.

## I. INTRODUCTION

Metal-oxide-semiconductor field-effect transistors (MOSFETs) based on 4H silicon carbide (4H-SiC) are applied to switching devices for high-power applications [1]. However, low conductivity at the channel interfaces between amorphous insulator and crystalline 4H-SiC remains a serious problem. The roughness at the interface is a possible cause of the low interface conductivity of SiC MOSFETs [2-5]. The surface-roughness-scattering model established for $SiO_2$/Si interfaces [6-10] has been partially used to analyze the mobility at $SiO_2$/4H-SiC interfaces [11-14]. For $SiO_2$/Si interfaces, the power spectral density of surface roughness was extracted from the boundary line between the two phases in the cross-sectional image obtained by a transmission electron microscope (TEM) [9], and mobility was calculated from the theoretical formula [6-8] and power spectral density. For SiC interfaces, however, power spectral density has not been determined from TEM images and has been treated as fitting parameters to reproduce measured electrical properties or mobilities [11-14].

It is a difficult task to draw the boundary line between the two phases. Goodnick *et al.* [9] defined interface boundary as "the last discernible lattice fringe corresponding to the periodicity of the Si," but they admitted that "this procedure is somewhat arbitrary at many points, as an abrupt change from crystalline Si to non-crystalline $SiO_2$ is not always apparent." They probably drew the boundary manually. Zhao *et al.* [15] took another approach: they defined the boundary based on the distinct darkness difference between the two phases. This approach is good in that the interface is determined uniquely and automatically. This approach, however, cannot be applied to SiC interfaces, because the TEM image of crystalline SiC has both bright and dark regions [2-4,16,17]. In addition, the existence of an interface transition layer has been ignored in previous reports.

A convolutional neural network (CNN) is a deep learning approach that has achieved great success in image classification [18-20]. In this study, we classified each point in a TEM image as amorphous insulator or crystalline 4H-SiC by using this approach and determined the interface boundary.

## II. DATASET PREPARATION

A TEM sample was prepared by depositing amorphous aluminum oxide on the 4-degree-tilted (0001) surface of crystalline 4H-SiC [21]. We chose the interface that was too rough to draw the boundary line manually. A cross-sectional image was obtained using a TEM (JEM-ARM200F, JEOL) at an acceleration voltage of 200 kV. The sample thickness was estimated to be approximately 70 nm. Figure 1 shows the cross-sectional TEM image, where the upper half is amorphous aluminum oxide (A) and lower half is crystalline 4H-SiC (C). Figure 2 shows the crystal structure of 4H-SiC.

The size of one pixel in Fig. 1 is 0.016 nm × 0.016 nm ($\Delta r$ = 0.016 nm). The training dataset consist of 2,000 small images that were randomly extracted from two distinct regions that are approximately 0.6-3 nm away from the interface; 1,000 images were extracted from A region in the TEM image (Fig. 1) and 1,000 from C. In the same way, 2,000 images for the test dataset were extracted from another two regions that are approximately 3-20 nm away from the interface. Every image in the training and test datasets does not include the interface and is assigned the correct label (A or C). The size of the extracted images is 29 pixels × 29 pixels (0.46 nm × 0.46 nm) and an example is shown in Fig. 3. This size was selected to include several lattice points (Fig. 2) in order that CNN can recognize the feature of the crystalline 4H-SiC. At every position (pixel) in the interface region, a 29 pixels × 29 pixels image centered on that position was extracted, which results in a dataset of 214,326 images (interface dataset). Images in the interface dataset are not assigned a correct label.

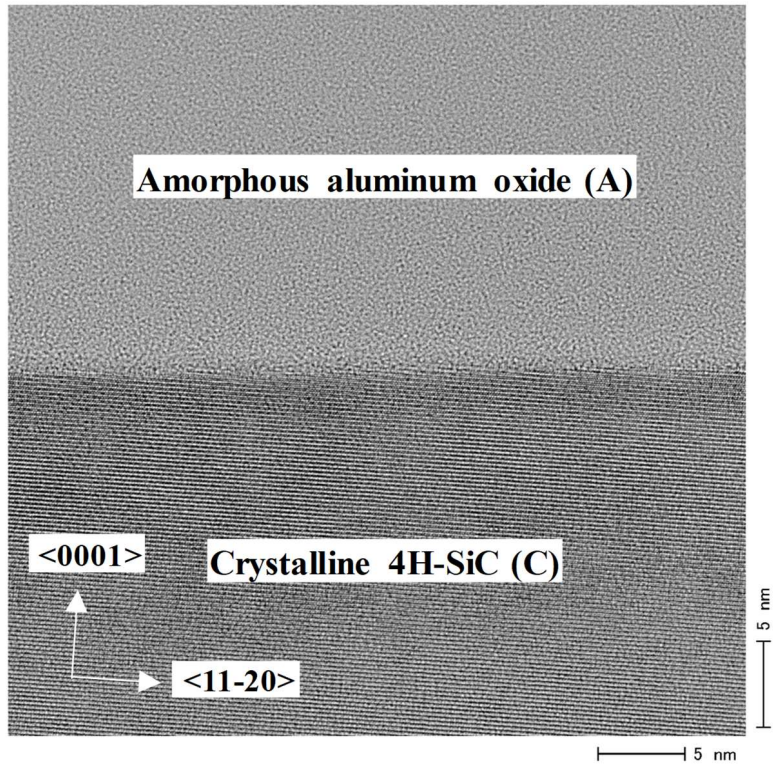

FIG. 1. Cross-sectional TEM image, where the upper half is amorphous aluminum oxide (A) and lower half is crystalline 4H-SiC (C).

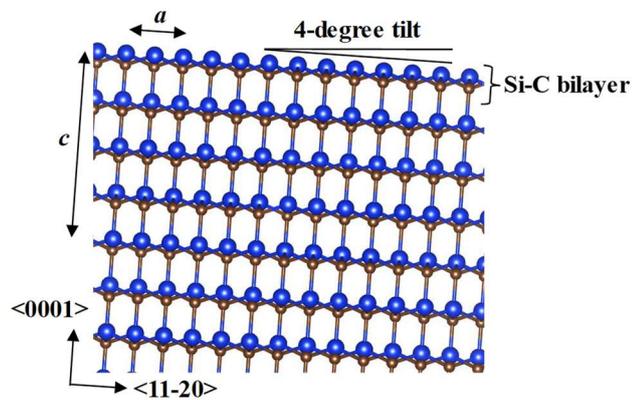

FIG. 2. Crystal structure of 4H-SiC, where blue and brown balls depict Si and C atoms, respectively. a = 0.307 nm and c = 1.005 nm are lattice constants of 4H-SiC.

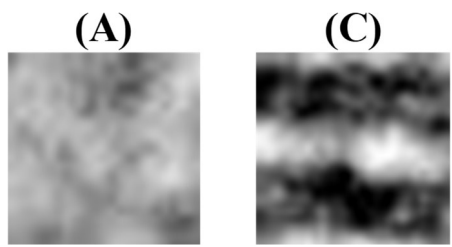

FIG. 3. An example of the images used for training: (A) amorphous aluminum oxide and (C) crystalline 4H-SiC. The size is 29 pixels × 29 pixels (0.46 nm × 0.46 nm).

## III. DEEP LEARNING MODELS

The software of neural network console (NNC) developed by Sony Network Communications Inc. was used in this study [22]. Deep learning models were trained using the training dataset to classify the extracted images as A or C, and the degree of learning was evaluated with the test dataset. Subsequently, every position (pixel) in the interface region were classified using the corresponding image in the interface dataset.

Figure 4(a) shows the convolutional neural network (CNN) model designed in this study based on Ref. [18,23]. Another model (FC) that consists of fully connected layers (affine layers) was also used as a reference and is shown in Fig. 4(b). The CNN used in this study is a modified version of CNN proposed in Ref. [18]. Figure 5(a) shows the configuration of our CNN. Main change is the introduction of Batch Normalization (BN) layers [23] after convolution layers. The BN layers prevent the distribution of internal variables from changing significantly and suppress overfitting. Adam [24] was used as an optimizer at $\alpha$ = 0.001. FC was also evaluated using the configuration shown in Figure 5(b). Figure 6 shows the learning progress with a batch size of 30, where the errors for both training and test datasets decreased with epoch. The CNN trained for 43 epochs was used to classify the interface dataset, and the FC model trained for 14 epochs was used. NNC default values were used for other hyperparameters.

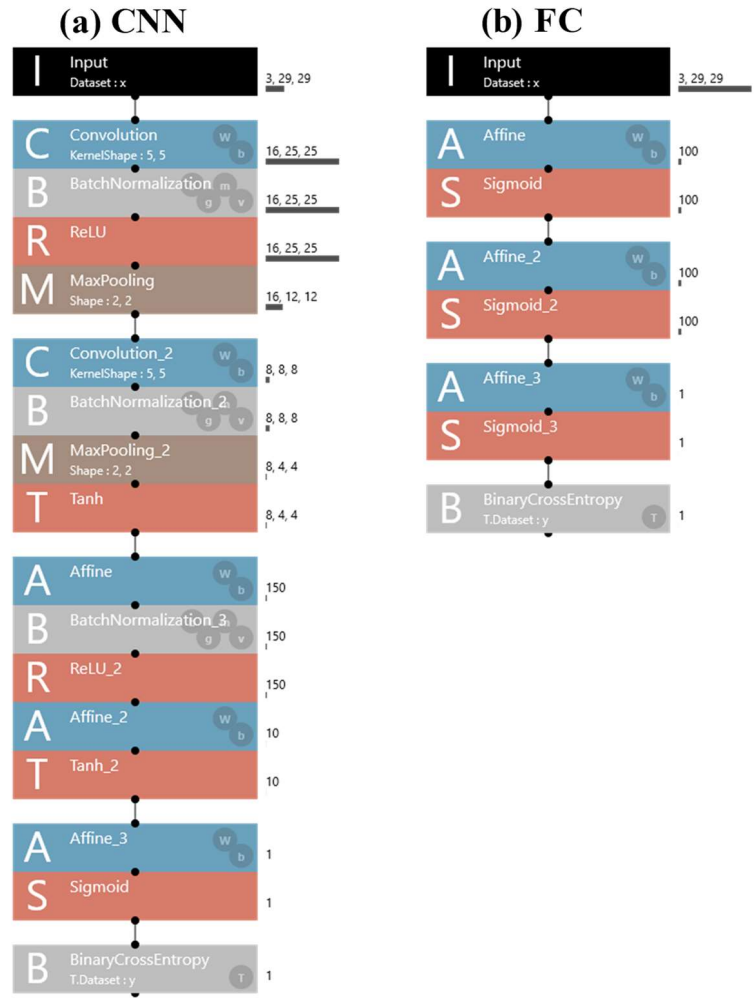

FIG. 4. (a) Convolutional neural network (CNN) model used for image classification, and (b) another model (FC) that consists of fully connected layers (affine layers).

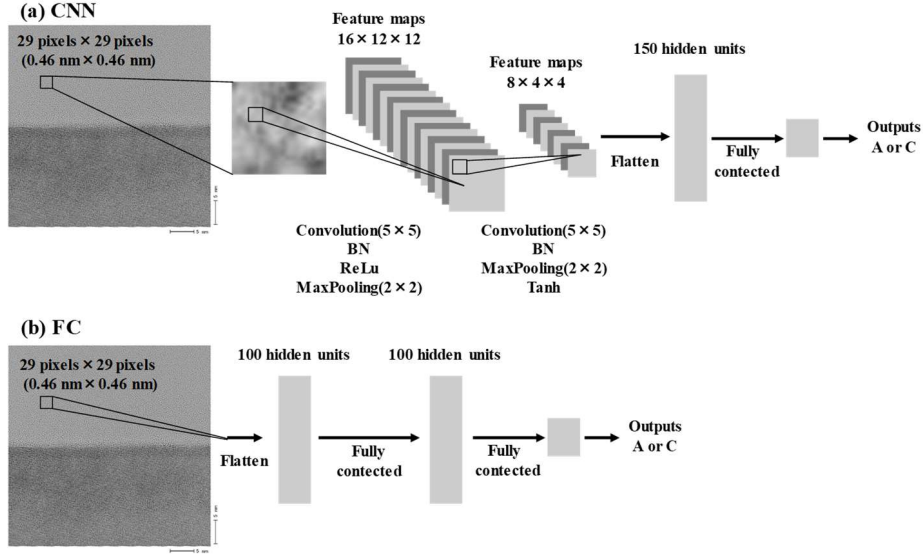

FIG. 5.   Configuration of (a) CNN and (b) FC used in this study. Outputs are whether the 29 pixels × by 29 pixels image is A or C.

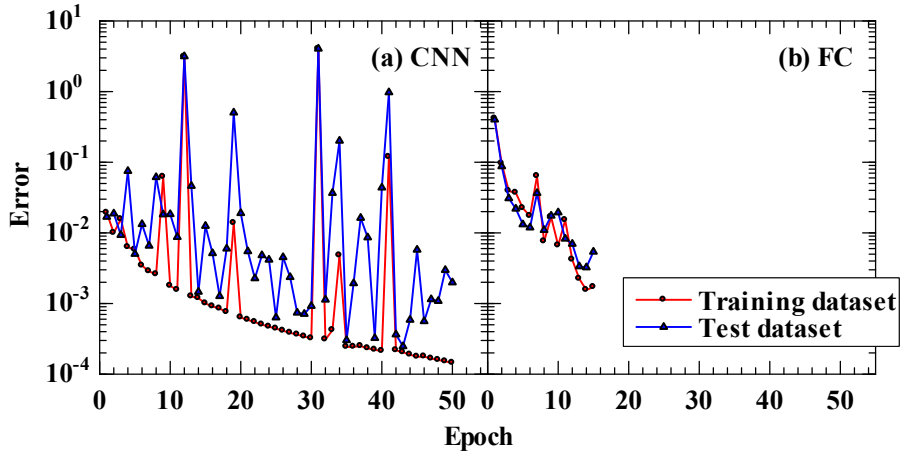

FIG. 6.   Learning progress for (a) CNN and (b) FC models.

## IV. RESULTS

Figure 7 shows the images of the interface region extracted from Fig. 1. Figures 7(b) and (c) show the classification results by FC and CNN models, respectively, where pixels classified as A (C) are colored red (yellow). The FC model could not classify the interface very well, e.g. it misjudged the lower part of the C region to be the A region. On the other hand, the CNN model, which is effective in image classification [18-20], classified the interface very well. For example, we humans can recognize a hollow at the position indicated by the arrow in Fig. 7. The shape of the hollow was well formed by the CNN model. Hereafter, only the CNN model will be used for further analysis.

It is possible to insert the third phase of an intermediate transition region, by assigning the pixels with intermediate output values (e.g. $0.001 < Y < 0.999$) to that region, which is shown as the red region in Fig. 7(d). The average thickness of the transition layer is calculated by dividing the area of the transition layer (red area in Fig. 7(d)) by length in the horizontal direction of the analyzed rectangular area. The calculated value was 0.29 nm, which is comparable to a Si-C bilayer in <0001> direction (0.25 nm). The ability to insert the intermediate transition region is an advantage of the deep learning approach that handwriting does not have.

For the purpose of evaluating the power spectral density of roughness, the boundary lines in Fig. 7(c) were modified to be a single-valued function shown in Fig. 7(e) by

$$z'(n_x) = \Delta r \sum_{n_z} Y(n_x, n_z),$$

$$z(n_x) = z'(n_x) - (An_x + B), \quad (1)$$

where $Y$ is the output value by the CNN model ($0 \leq Y \leq 1$), $n_x$ (1−2,646) and $n_z$ (1−81) are the serial numbers of pixels in the horizontal and vertical directions, respectively. The least squares regression line ($An_x+B$) is subtracted from $z'(n_x)$. The calculated root-mean-square (standard deviation) of $z(n_x)$ was 0.14 nm.

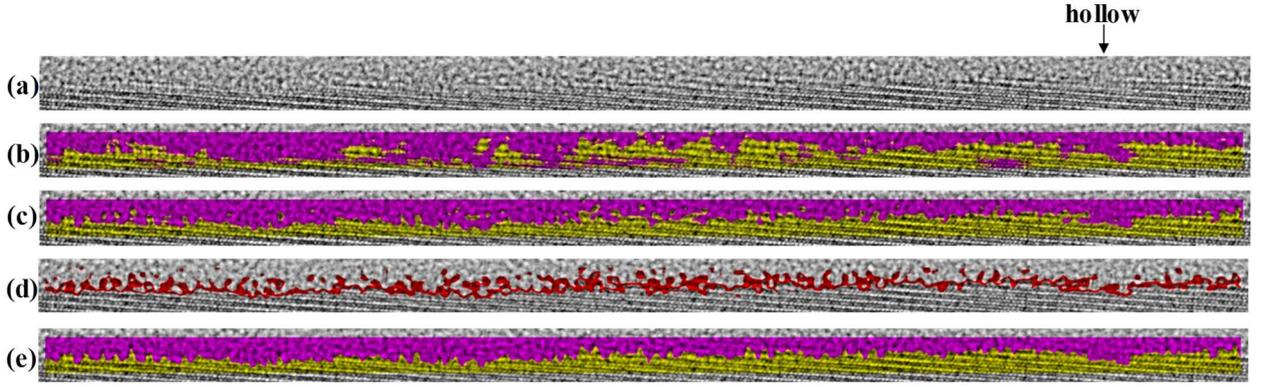

FIG. 7. Images of the interface region extract from Fig. 1. (a) Image before classification. Classification results by (b) FC and (c) CNN models shown in Figs. 4 and 5, where pixels classified as A (C) are colored red (yellow). (d) Intermediate transition region that was classified under the condition of output values being 0.001 < Y < 0.999 by CNN model and is colored red. (e) Boundary line that is a single-valued function, which was calculated by Eq. 1 from the results by CNN model. The size of classified rectangular is 2,646 pixels × 81 pixels (42 nm × 1.3 nm).

Figure 8 shows the power spectral density calculated by using discrete Fourier transform as

$$S(n_q) = \frac{\Delta r}{N_x} \left| \sum_{n_x=1}^{N_x} z(n_x) \exp\left(\frac{2\pi i (n_x-1)(n_q-1)}{N_x}\right) \right|^2, \quad (2)$$

$$x = (n_x - 1)\Delta r, \quad q = (n_q - 1)\Delta q, \quad \Delta q = \frac{2\pi}{\Delta r N_x} = 0.15 \text{ nm}^{-1},$$

where $N_x$ = 2,646 is the total pixel number in the horizontal direction. We can recognize two peaks at $q$ = 0.6 nm$^{-1}$ and 11 nm$^{-1}$. The peak at 0.6 nm$^{-1}$ corresponds to the periodic step-and-terrace structure on the 4-degree-tilted surface with steps consisting of two Si-C bilayers ($q$ = 0.87 nm$^{-1}$). In other words, the CNN model revealed step bunching with two bilayers. The peak at 11 nm$^{-1}$ indicates that there is the fluctuation in the order of several atoms. This fluctuation seems to stem from the intermediate region where the A and C regions overlap vertically on the paper surface. It is difficult for human eyes to classify these ambiguous regions, but the CNN model can classify them. The size of the extracted images used in the classification process was 0.46 nm ($q$ = 14 nm$^{-1}$), then structures smaller than this order cannot be detected. Thus, it is reasonable that the power spectral density drastically decreases for $q$ > 14 nm$^{-1}$ as shown in Fig. 8.

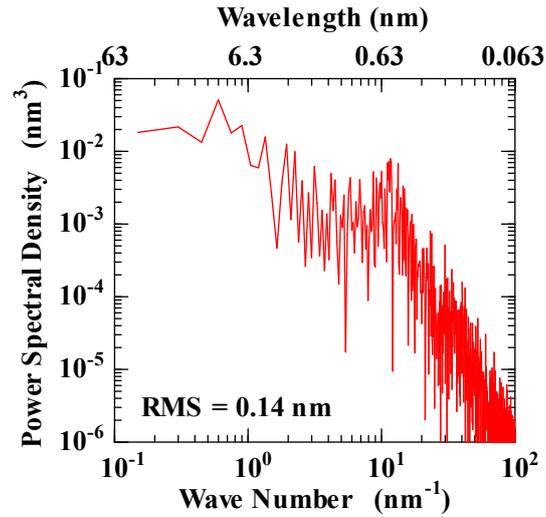

FIG. 8.  Power spectral density of interface roughness shown in Fig. 7(e) calculated by Eq. (2).

Although classification results depend on the CNN model and its hyperparameters, we expect that it can reveal the morphological differences between different samples by using the same model and hyperparameters. We would like to evaluate interfaces formed under various conditions in our future studies.

**V. CONCLUSION**

We drew the boundary line between two phases by using convolutional neural network (CNN). We show that the CNN model can recognize the interface accurately, even when the interface is too rough to draw the boundary line manually.

**SUPPLEMENT**

Figures below are Figs. 1, 3(A), 3(C), and 7(a)-7(e) that directly converted to PDF.

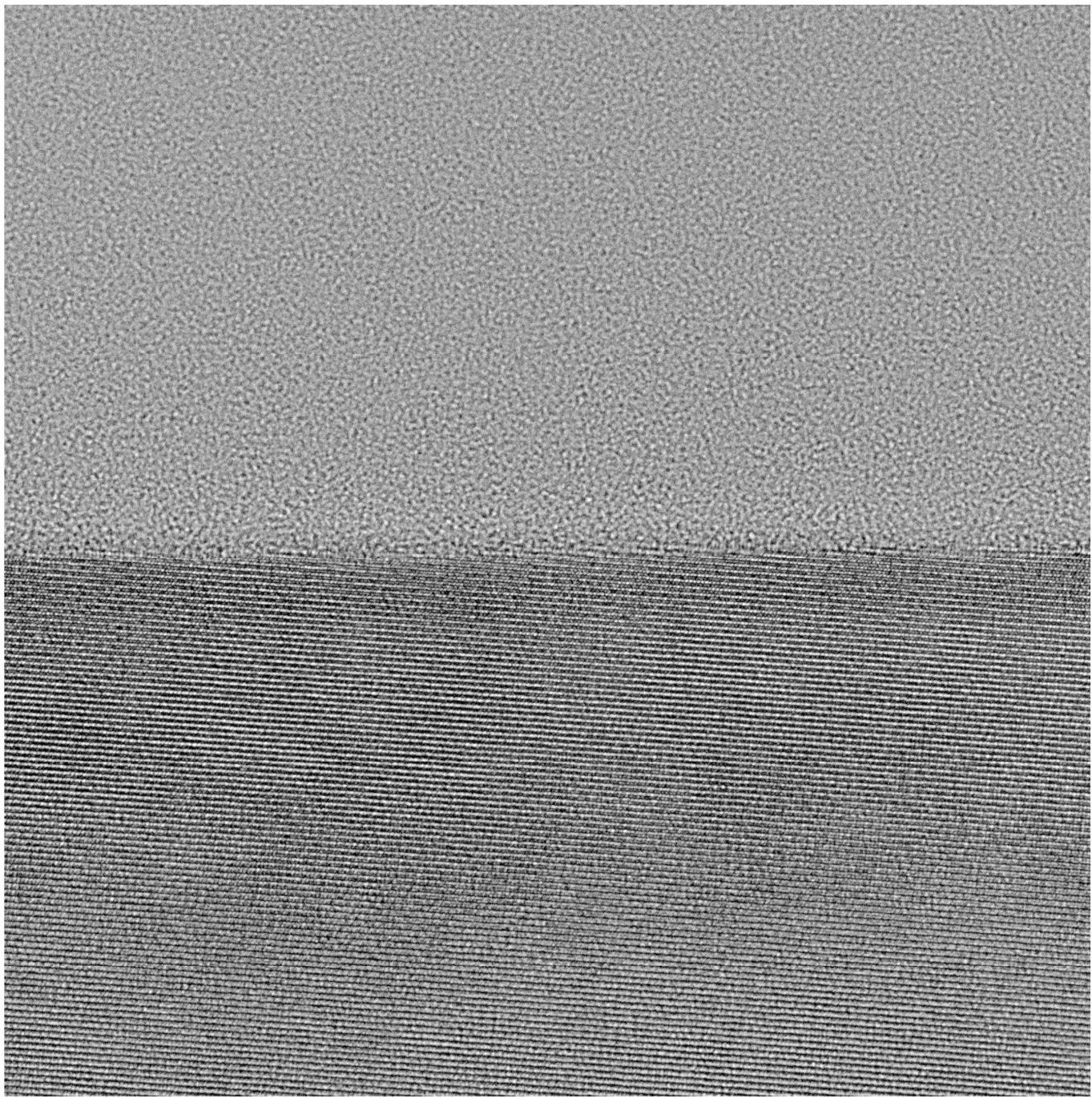

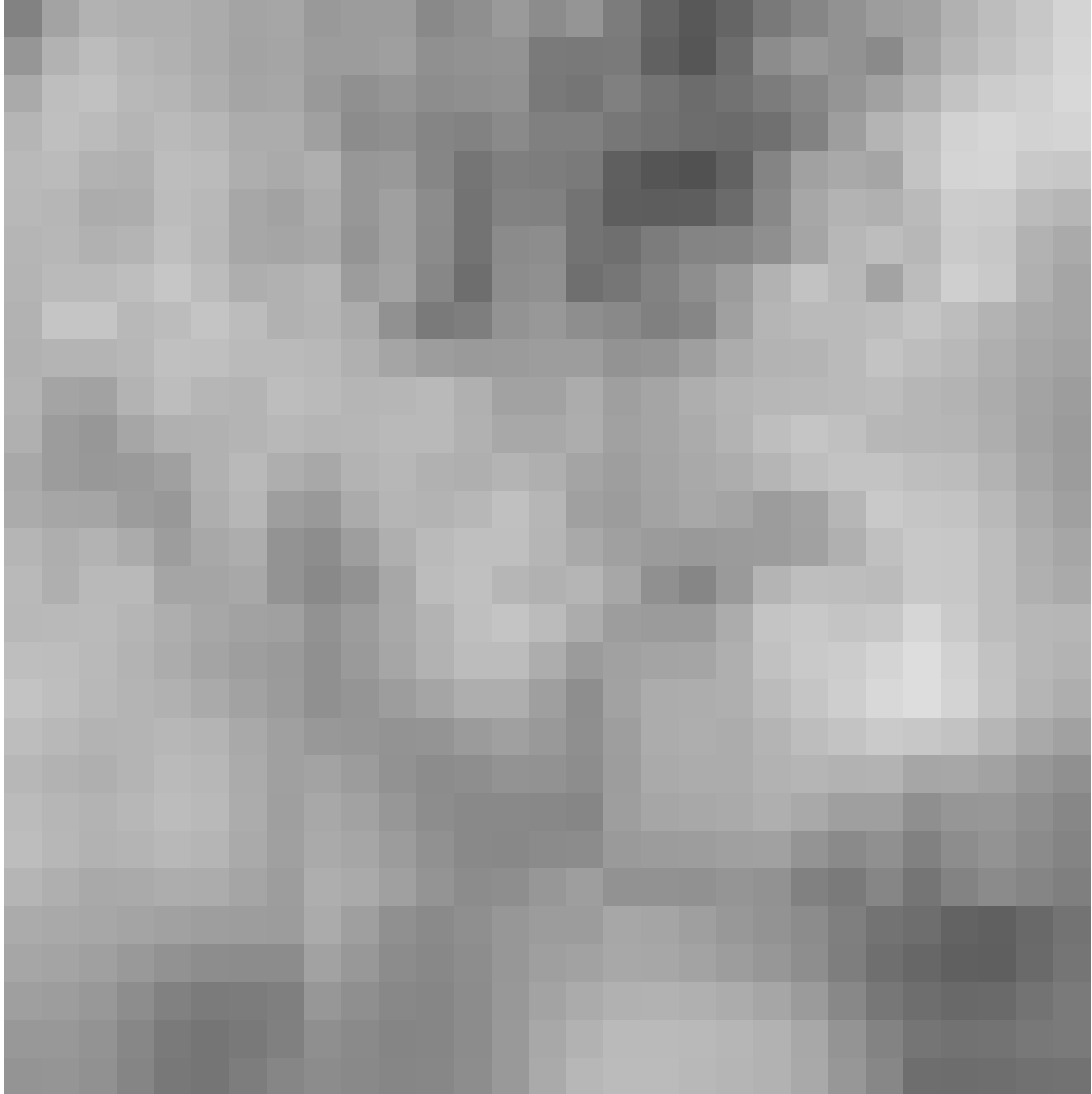

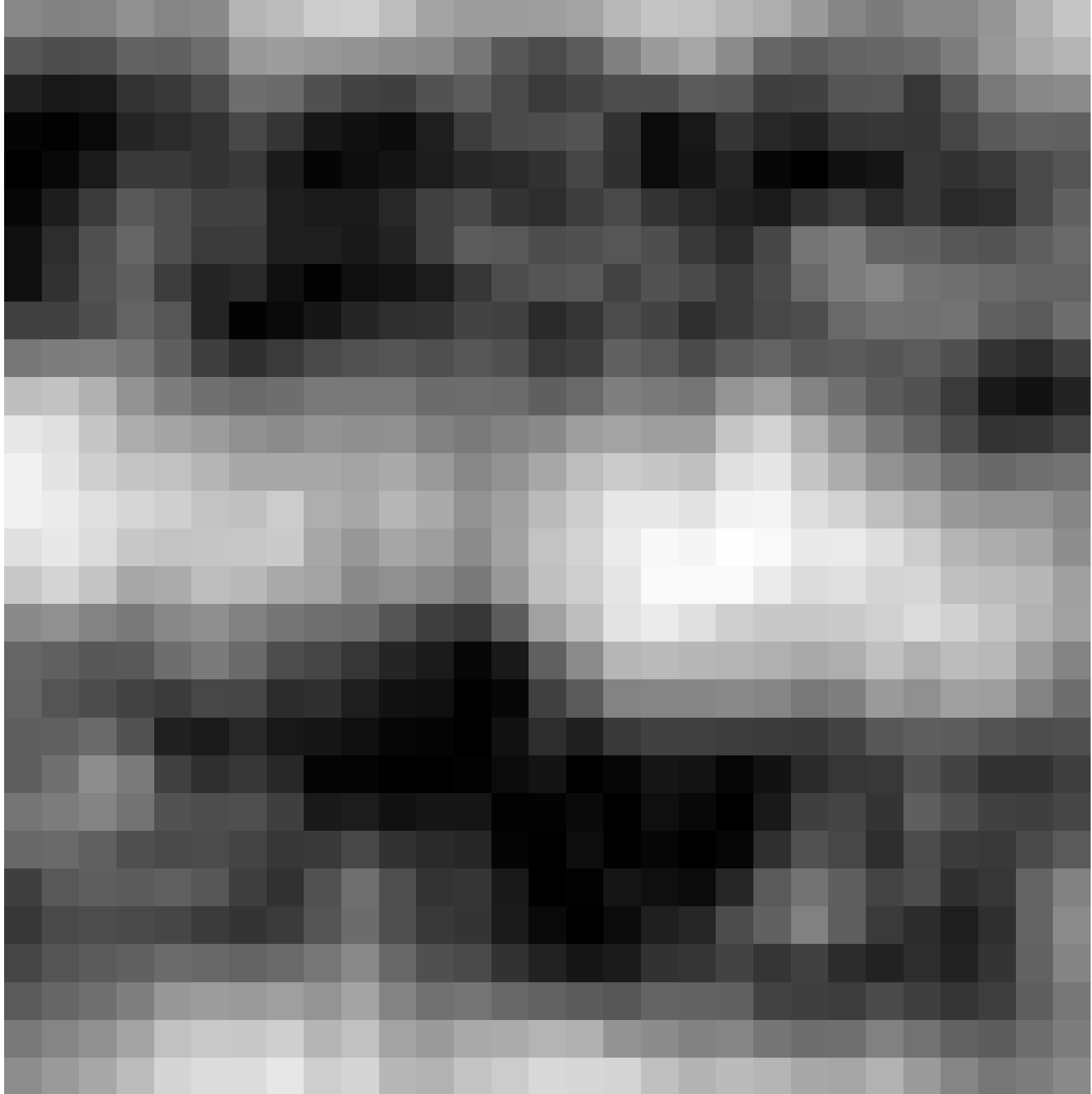

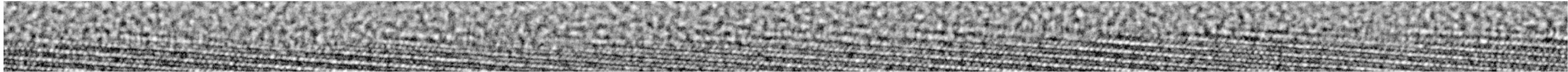

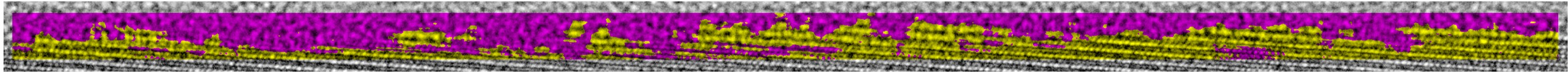

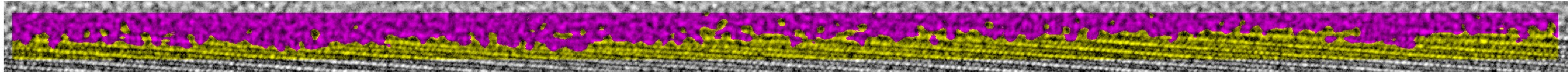

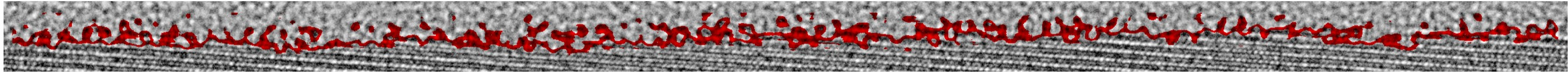

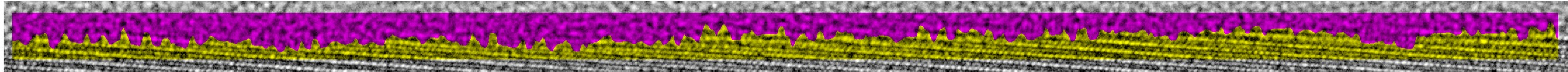